\documentclass[12pt]{article}

\pdfoutput=1

\usepackage{cite}

\usepackage{amssymb,latexsym}

\usepackage{epstopdf}

\usepackage{graphics,epsfig}

\usepackage{color}

\usepackage{amsmath,amsfonts}

\usepackage{mathrsfs}

\DeclareMathAlphabet{\mathpzc}{OT1}{pzc}{m}{it}

\usepackage{tikz}
\usetikzlibrary{arrows,shapes}
\usetikzlibrary{trees}
\usetikzlibrary{matrix,arrows} 				
\usetikzlibrary{positioning}				
\usetikzlibrary{calc,through}				
\usetikzlibrary{decorations.pathreplacing}  
\usepackage{pgffor}							

\usetikzlibrary{decorations.pathmorphing}	
\usetikzlibrary{decorations.markings}
\tikzset{
    vector/.style={decorate, decoration={snake}, draw},
	provector/.style={decorate, decoration={snake,amplitude=2.5pt}, draw},
	antivector/.style={decorate, decoration={snake,amplitude=-2.5pt}, draw},
    fermion/.style={draw=black, postaction={decorate},
        decoration={markings,mark=at position .55 with {\arrow[draw=black]{>}}}},
    fermionbar/.style={draw=black, postaction={decorate},
        decoration={markings,mark=at position .55 with {\arrow[draw=black]{<}}}},
    fermionnoarrow/.style={draw=black},
    gluon/.style={decorate, draw=black,
        decoration={coil,amplitude=4pt, segment length=5pt}},
    scalar/.style={dashed,draw=black, postaction={decorate},
        decoration={markings,mark=at position .55 with {\arrow[draw=black]{>}}}},
    scalarbar/.style={dashed,draw=black, postaction={decorate},
        decoration={markings,mark=at position .55 with {\arrow[draw=black]{<}}}},
    scalarnoarrow/.style={dashed,draw=black},
    electron/.style={draw=black, postaction={decorate},
        decoration={markings,mark=at position .55 with {\arrow[draw=black]{>}}}},
	bigvector/.style={decorate, decoration={snake,amplitude=4pt}, draw},
}

\tikzstyle{block} = [draw, rectangle, 
    minimum height=3em, minimum width=6em]

\let\a=\alpha \let\b=\beta \let\g=\gamma \let\d=\delta \let\e=\epsilon
\let\z=\zeta  \let\th=\theta  \let\k=\kappa
\let\l=\lambda \let\m=\mu \let\n=\nu \let\x=\xi \let\p=\pi 
\let\s=\sigma   \let\f=\phi  
 
        \let\Th=\Theta \let\L=\Lambda
\let\X=\Xi  \let\S=\Sigma  \let\Y=\Psi
 \let\W=\Omega
\let\la=\label  
  
\def\nn{\nonumber} \def\bd{\begin{document}} \def\ed{\end{document}}
\def\ds{\documentstyle} \let\fr=\frac \let\bl=\bigl \let\br=\bigr
\let\Br=\Bigr \let\Bl=\Bigl
\let\bm=\bibitem
\let\na=\nabla
\def\tU{{\widetilde U}}
\let\pa=\partial \let\ov=\overline
\def\ie{{\it i.e.\ }}
\newcommand{\be}{\begin{equation}}
\newcommand{\ee}{\end{equation}}
\def\ba{\begin{array}}
\def\ea{\end{array}}
\def\ft#1#2{{\textstyle{{\scriptstyle #1}\over {\scriptstyle #2}}}}
\def\fft#1#2{{#1 \over #2}}
\def\F#1#2{{ F_{#1}^{(#2)} }}
\def\cF#1#2{{ {\cal F}_{#1}^{(#2)} }}

\def\R{{\bf R}}
\def\sst#1{{\scriptscriptstyle #1}}
\def\oneone{\rlap 1\mkern4mu{\rm l}}
\def\e7{E_{7(+7)}}
\def\td{\tilde}
\def\wtd{\widetilde}
\def\im{{\rm i}}
\def\bog{Bogomol'nyi\ }
\newcommand{\ho}[1]{$\, ^{#1}$}
\newcommand{\hoch}[1]{$\, ^{#1}$}
\newcommand{\bea}{\begin{eqnarray}}
\newcommand{\eea}{\end{eqnarray}}
\newcommand{\ra}{\rightarrow}
\newcommand{\lra}{\longrightarrow}
\newcommand{\Lra}{\Leftrightarrow}
\newcommand{\ap}{\alpha^\prime}
\newcommand{\bp}{\tilde \beta^\prime}
\newcommand{\cB}{{\cal B}}
\newcommand{\cO}{{\cal O}}
\newcommand{\vecx}{\vec{x}}
\newcommand{\vecy}{\vec{y}}
\newcommand{\vecp}{\vec{p}}
\newcommand{\vecq}{\vec{q}}
\newcommand{\tr}{{\rm tr} }
\newcommand{\Tr}{{\rm Tr} }
\newcommand{\NP}{Nucl. Phys. }

\newcommand{\cL}{{\cal L}}
\newcommand{\cA}{{\cal A}}
\newcommand{\cT}{{\cal T}}
\newcommand{\cR}{{\cal R}}
\newcommand{\cD}{{\cal D}}
\newcommand{\cH}{{\cal H}}

\def\Cb{\bar{C}}

\def\sst#1{{\scriptscriptstyle #1}}
\def\0{{\sst{(0)}}}
\def\1{{\sst{(1)}}}
\def\2{{\sst{(2)}}}
\def\3{{\sst{(3)}}}
\def\4{{\sst{(4)}}}
\def\5{{\sst{(5)}}}
\def\6{{\sst{(6)}}}
\def\7{{\sst{(7)}}}
\def\8{{\sst{(8)}}}
\def\9{{\sst{(9)}}}
\def\p{{\sst{(p)}}}
\def\q{{\sst{(q)}}}
\def\ve{\varepsilon}
\def\vf{\varphi}
\def\F{\Phi}
\def\wg{\wedge}

\def\thb{\bar{\theta}}
\def\Thb{\bar{\Theta}}
\def\barp{\bar{p}}
\def\barq{\bar{q}}
\def\barc{\bar{c}}
\def\bard{\bar{d}}
\def\e{\epsilon}

\def \bi{\bibitem}
\def \la {\label}

\def \l {\lambda}
\def\foot{\footnote}
\def \tl  {{\tilde \l}}
\def \sql {{\sqrt \l}}
\def \adss {$AdS_5 \times S^5$\ }
\newcommand{\rf}[1]{(\ref{#1})}
\def \ov {\over}

\def\th{\theta}
\def\Th{\Theta}
\def\vth{\vartheta}
\def\btheta{{\bar\theta}}
\def\ttheta{{{\tilde\theta}}}
\def\bttheta{{{\bar\ttheta}}}
\def\vth{\vartheta}

\def\ra{\rightarrow}
\def\N{\nabla}
\def\F{{\cal F}}
\def\uM{\underline{M}}
\def\uA{\underline{A}}
\def\uN{\underline{N}}
\def\uP{\underline{P}}
\def\ua{\underline{a}}
\def\ub{\underline{b}}
\def\uc{\underline{c}}
\def\ud{\underline{d}}
\def\ue{\underline{e}}
\def\uf{\underline{f}}
\def\ui{\underline{i}}
\def\uj{\underline{j}}
\def\uk{\underline{k}}
\def\ul{\underline{l}}
\def\ual{\underline{\alpha}}
\def\ube{\underline{\beta}}
\def\um{\underline{m}}
\def\un{\underline{n}}
\def\up{\underline{p}}
\def\uq{\underline{q}}
\def\ur{\underline{r}}
\def\us{\underline{s}}
\def\umu{\underline{\mu}}
\def\unu{\underline{\nu}}
\def\ula{\underline{\l}}
\def\uka{\underline{\k}}
\def\usi{\underline{\s}}
\def\urh{\underline{\r}}
\def\cc{\circ}
\def\eqv{\equiv}

\def\ni{\noindent}

\def\Ep{E^{{}^{(+)}}}
\def\Em{E^{{}^{(-)}}}

\def\Mp{M^{{}^{(+)}}}
\def\Mm{M^{{}^{(-)}}}

\def \ha{{1\ov 2}}

\def\r{\rho}

\def\Y{{\rm Y}}
\def\X{{\rm X}}
\def\tY{\tilde{\rm Y}}
\def\tX{\tilde{\rm X}}
\def\dY{\dot{\rm Y}}
\def\dX{\dot{\rm X}}

\def \J {\mathcal{J}}
\def \del {\partial}

\def\dF{\dot{F}}
\def\dG{\dot{G}}
\def\df{\dot{f}}
\def \E {{\cal E}}
\def \S {{\cal S}}
\def \J {{\cal J}}

\def\ms{\mathcal{S}}
\def\mj{\mathcal{J}}
\def\soj{\fr{\ms}{\mj}}
\def \R {{\bf R}}
\def \om {\omega}
\def \bE {\bar E}
\def \x {{\cal X}}

\def \bi{\bibitem}
\def \la {\label}

\def \l {\lambda}
\def\foot{\footnote}
\def \tl  {{\tilde \l}}
\def \sql {{\sqrt \l}}
\def \adss {$AdS_5 \times S^5$\ }
\def \ov {\over}

\def \varpi {{\rm w}}

\def\thb{\bar{\theta}}
\def\Thb{\bar{\Theta}}
\def\mb{\bar{\m}}
\def\ab{\bar{\a}}
\def\zb{\bar{z}}
\def\psib{\bar{\psi}}
\def\barp{\bar{p}}
\def\barq{\bar{q}}
\def\barc{\bar{c}}
\def\bard{\bar{d}}
\def\e{\epsilon}
\def\wb{\bar{w}}
\def\lb{\bar{\l}}
\def\Jb{\bar{J}}
\def\Nb{\bar{N}}
\def\Zb{\bar{Z}}
\def\pab{\bar{\pa}}

\def\At{\tilde{A}}
\def\Bt{\tilde{B}}
\def\Ct{\tilde{C}}
\def\Dt{\tilde{D}}
\def\Et{\tilde{E}}
\def\Ft{\tilde{F}}
\def\Gt{\tilde{G}}
\def\Ht{\tilde{H}}
\def\Kt{\tilde{K}}
\def\Mt{\tilde{M}}
\def\Nt{\tilde{N}}
\def\Rt{\tilde{R}}
\def\at{\tilde{a}}
\def\bt{\tilde{b}}
\def\ct{\tilde{c}}
\def\dt{\tilde{d}}
\def\et{\tilde{e}}
\def\ft{\tilde{f}}
\def \ztt{\tilde{\z}}
\def \zetat{\tilde{\zeta}}
\def\htil{\tilde{h}}
\def\gt{\tilde{g}}
\def\nt{\tilde{n}}
\def\mut{\tilde{\mu}}
\def\nut{\tilde{\nu}}
\def\pht{\tilde{\f}}
\def\Phit{\tilde{\Phi}}
\def\vft{\tilde{\vf}}

\def\rht{\tilde{\rho}}

\def\asth{\hat{*}}
\def\phh{\hat{\phi}}

\def\bA{{\bf A}}

\def\ola{\overleftarrow}
\def\ora{\overrightarrow}
\def\alt{\tilde{\a}}

\def\eh{\hat{e}}
\def\eph{\hat{\e}}
\def\ph{\hat{p}}
\def\alh{\hat{\a}}
\def\beh{\hat{\b}}
\def\gah{\hat{\g}}
\def\Fh{\hat{F}}
\def\muh{\hat{\m}}
\def\nuh{\hat{\n}}
\def\thh{\hat{\th}}
\def\rhh{\hat{\r}}
\def\dh{\hat{d}}
\def\ih{\hat{i}}
\def\jh{\hat{j}}
\def\hh{\hat{h}}
\def\nh{\hat{n}}
\def\gh{\hat{g}}
\def\kh{\hat{k}}
\def\deh{\hat{\d}}
\def\wh{\hat{w}}
\def\lah{\hat{\l}}
\def\Ah{\hat{A}}
\def\Kh{\hat{K}}
\def\Nh{\hat{N}}
\def\Rh{\hat{R}}
\def\Ch{\hat{C}}
\def\Omh{\hat{\Omega}}

\def\xh{\hat{x}}

\def\ps{\rlap{\, /}\;\,p }
\def\ks{\rlap{\, /}\;\,k }

\def\gym{g_{YM}}

\def\adot{\dot{a}}
\def\bdot{\dot{b}}
\def\bpa{\bar{\pa}}

\def\pr{\prime}
\def\ssk{\medskip}
\def\clb{\color{blue}}
\def\clr{\color{red}}
\def\clg{\color{green}}

\def\bfA{{\bf A}}
\def\bfB{{\bf B}}
\def\bfK{{\bf K}}
\def\bfU{{\bf U}}
\def\bfX{{\bf X}}
\def\bfY{{\bf Y}}
\def\bfZ{{\bf Z}}
\def\bfg{{\bf g}}
\def\bfn{{\bf n}}

\def\bsk{\bigskip}
\def\ssk{\medskip}

\def\Ec{{\cal E}}

\begin{document}

\overfullrule=0pt
\parskip=2pt
\parindent=12pt
\headheight=0in \headsep=0in \topmargin=0in
\oddsidemargin=0in

\vspace{ -3cm}
\thispagestyle{empty}

 \vspace{0.1cm}

\setcounter{equation}{0}
\setcounter{footnote}{0}
\setcounter{section}{0}

\begin{center}

{\Large\bf  Quantum-induced trans-Planckian energy near horizon}

\vskip 0.8cm

\vspace{0.5cm}
A. J. Nurmagambetov$\,^{\spadesuit}$\let\thefootnote\relax\footnotetext{$^{\spadesuit}$ Also at {\it Karazin Kharkov National University, 4 Svobody Sq., Kharkov, UA 61022} \& {\it Usikov Institute for Radiophysics and Electronics, 12 Proskura St., Kharkov, UA 61085}. } 
and I. Y. Park{$^\dagger$}
\\

\vspace{0.3cm}

$^{\spadesuit}$
{\it Akhiezer Institute for Theoretical Physics of
NSC KIPT,\\
1 Akademicheskaya St., Kharkov, \\ UA 61108 Ukraine \\
ajn@kipt.kharkov.ua
}

\vspace{0.3cm}
{\it {}{$^\dagger$}Department of Applied Mathematics,
Philander Smith College 
                               \\
Little Rock, AR 72223, USA \\
inyongpark05@gmail.com
}

 \vspace{.5cm}

\end{center}

 \vspace{0.1cm}

\begin{abstract}

We study the loop effects on the geometry  and boundary conditions of several black hole spacetimes one of which is time-dependent and analyze the energy measured by an infalling observer near their horizons.
The finding in the previous works that the loop effects can be drastic is reinforced: they play an important role in the boundary conditions and non-perturbative geometry deformation. 
One of the channels through which the quantum gravitational effects enter is generation of the cosmological constant. The cosmological constant feeds part of the time-dependence of a solution.  
We obtain a trans-Planckian energy in the time-dependent case. The importance of time-dependence for the trans-Planckian energy and black hole information is discussed.

\end{abstract}
\newpage

\section{Introduction}

The remarkable developments of astrophysical observations such as the detection of the gravitational wave \cite{Abbott:2016blz} or  the Event Horizon Telescope project \cite{Doeleman:2009te} could offer valuable guidance to the correct formulation of the theory of quantum gravity (see, e.g.,\cite{Burgess:2006bm,Donoghue:2015hwa,Haggard:2016ibp,Carlip:2017eud,Calcagni:2017shx} for various reviews from theoretical and observational points of views). Black holes provide an optimal arena for studying quantum gravitational physics: they are mathematically simple at the classical level yet require quantization for a complete and proper understanding. In particular, the black hole information problem\footnote{See \cite{Page:1993up,Hooft:2016vug,Guo:2017jmi} and references therein for reviews; recent works include \cite{Dvali:2015aja,Kawai:2017txu,Ho:2017vgi,Lust:2017gez,Chu:2018tzu,Unruh:2017uaw}.}   
poses a challenge that will, once surmounted, take us to the next chapter of understanding astrophysical black holes at a more fundamental level.

Sometime ago interest in the black hole information problem was renewed by the Firewall argument \cite{Almheiri:2012rt,Braunstein:2009my} followed by various debates.
One of the facts brought home - perhaps more systematically than ever - by the Firewall observation is that our understanding of black holes and gravitational physics as a whole is as yet incomplete.
The Firewall proposal has challenged, among other things, the conventional view that a free-falling observer would not experience anything out of the ordinary when passing through the horizon: the observer should encounter trans-Planckian energy radiation. We have recently proposed in \cite{Park:2013rm} that quantum gravitational effects should be responsible for the production of high energy radiation, and ultimately, may well hold the key to the information puzzle.  
 Although the study of the information problem has a long history, two critical ingredients that could have led to a firmer grip of the problem had not, in the past, been quantitatively taken into account in the way they have been in our recent and present works. They are quantum gravitational effects and non-Dirichlet boundary conditions. We will examine them in detail in the main body by taking three cases of black hole spacetimes and continuing the endeavor initiated in \cite{Park:2013rm}, \cite{Park:2014mba} and \cite{Park:2017dib}. In particular, we analyze the quantum gravity-induced energy measured by an infalling observer near each horizon.

In the past, the apparent loss of information led to suspicion on a certain unknown information bleaching mechanism and the potential relevance of the quantum gravitational effects on the information problem was considered in the literature; see the review\cite{Page:1993up}.
However, the idea was not pursued at a quantitative level (presumably because of the difficulty of seeing what process could possibly be responsible for such bleaching). In \cite{Park:2013rm,Park:2014mba,Park:2017dib,Park:2017wiw}, we have unraveled a potential mechanism: one facet of the quantum gravitational effects should be as an information bleaching process.

The issue of the boundary conditions in gravitational physics seems profound. (See, e.g., \cite{Park:1998yw,Parattu:2015gga,Krishnan:2016dgy} for progress in boundary conditions and dynamics.) It awaits a more complete and comprehensive treatment in the context of quantization. A meaningful observation on the boundary conditions as a crucial component of the Hilbert space has recently been put forth in loop quantum gravity works \cite{Donnelly:2016auv} and \cite{Freidel:2016bxd}. Although the widely-used Dirichlet boundary conditions have been successful in non-gravitational areas, narrowing down to the configurations with these boundary conditions in a gravity theory results in wipe-out of much of the information as demonstrated in the recent works \cite{Park:2013rm,Park:2017wiw,Park:2016fxc,Park:2016vam}. The surprising fact that the Dirichlet boundary condition is at odds with the information of the system seems quite sophisticated, and must be due to the fact that the physical states of a gravitational system happen to have their support at the boundary hypersurface, the holographic screen, which in turn has its origin in the large amount of the gauge symmetry of a gravitational system \cite{Park:2014tia}.

The aforementioned two ingredients are not independent but intricately intertwined. As we will show, there exists a close influence of the quantum effects on the boundary conditions and geometry, especially the time-dependent one. The influences among these entities are mutual though we will take the quantum effect-centered view. The influence of quantum effects on the geometry is quite natural \cite{Park:2013rm}\cite{Park:2017dib}. The way the boundary conditions figure into the mutual relations has just been recognized \cite{Park:2016fxc,Park:2016vam}. One of the focuses of the present work is the manner in which the quantum effects and boundary conditions feed the time-dependence of a solution.

There are several routes to probing the perturbative loop effects on the geometry and physics. One of the approaches that can be taken with a reasonable amount of calculations is to study the deformation of the geometry analyzed through the 1PI effective action \cite{Davies:1976ei,Mukhanov:1994ax}. (See, e.g., \cite{Buchbinder} for a review of the 1PI effective action in the gravity context.) A more effortful direction would be the one based on a wave-packet - more in the conclusion. If one additionally works out the geodesics, it is straightforward to calculate the energy measured by an infalling observer, although the algebra involved is usually heavy. In \cite{Park:2017dib}, one-loop correction terms in the 1PI action were examined to see whether they would lead to a trans-Planckian energy when evaluated in a time-independent background, a Schwarzschild-dS background. They did not.  The analysis was then repeated for the time-dependent quantum-level background; there it was revealed that they do yield a trans-Planckian energy.\footnote{In \cite{Kawai:2017txu} (see also the earlier related work \cite{Kawai:2013mda}), it was observed in a time-dependent setup that the quantum stress-energy tensor inside the black hole reaches a near-Planckian value.} One of the key lessons learned through those (and the present) analyses is that there exist circumstances, such as when nonperturbative physics are relevant, where the quantum gravitational effects cannot be set aside as small. In the conclusion, we will argue that such circumstances must be quite common rather than exceptional. 

Two questions may arize. Firstly, in the case of the time-independent solution, could it be the high degree of symmetry of the solution that suppresses the trans-Planckian behavior? The time-independent background with less symmetry should be worth examining. Secondly, one may wonder whether or not the fact that quantum effects feed a time-dependent solution would persist in other cases. Put differently, how generic is the existence of time-dependent solutions fed by quantum effects? These questions will be addressed in the main body.

\vspace{.3in}
The paper is organized as follows. In section 2 we start with some of the salient features of quantization of gravity recently proposed in \cite{Park:2014tia,Park:2014noa,Park:2016zgt}. The quantization procedure {\em generically} leads to a quantum-corrected and/or -generated cosmological constant that in turn has a significant impact on solution generation: its presence contributes to a qualitatively  different - in the sense that the solution is time-dependent - solution. This is a non-perturbative effect, though the 1PI action is obtained perturbatively, through the back reaction of the metric and matter fields. Certain conceptual as well as technical aspects of the quantization procedure are essential: they not only provide the foundation on which the subsequent analysis is laid but also reveal some crucial aspects of the cosmological constant. 
The main theme of the present computation is the energy measured by an infalling observer. 
Because the tasks involved require intensive analyses, we illustrate the procedure with a simpler background, a Schwarzschild-Melvin solution. Then we consider another more complex stationary background; it is the recently constructed generalization of Schwarzschild-Melvin solution \cite{Preston:2006ze}. 
After recalling the findings in the previous works of \cite{Park:2016fxc}, \cite{Park:2016vam} and \cite{Park:2017dib}, we consider a time-dependent black hole spacetime in section 3. It is an extension of the time-dependent black hole solution previously obtained in \cite{Murata:2010dx}. The same kind of the trend as observed in \cite{Park:2017dib} is also observed: whereas the classical terms do not give the Firewall energy, the quantum effects do lead to a trans-Planckian energy. In the conclusion, we end with further remarks and future directions.

\section{Time-independent cases}

In this section we demonstrate the steps of the energy computation with time-independent black holes. 
For calculating the one-loop-corrected energy measured by an infalling observer, one needs to obtain the one-loop geodesic as well as the stress-energy tensor in the background under consideration. This has been carried out in \cite{Park:2017dib} for the case of a Schwarzschild-dS background: although the Schwarzschild-dS background itself did not, at least at one-loop, lead to a trans-Planckian energy, its time-dependent quantum extension did lead to a trans-Planckian energy. Here we are to look into the possibility that a time-independent solution with less symmetry might lead to a trans-Planckian energy. The result again yields a negative answer up to a certain subtlety that we will discuss below (and we turn to a time-dependent case and find an affirmative answer in section 3).

Let us consider the Einstein-Maxwell action,
\bea
S=\int \sqrt{-g}\;\Big(\fr1{\k^2}R-\fr14 F_{\m\n}^2 \Big)
\eea
The metric field equation is
\bea
G_{\m\n}=8\pi G{T}_{\m\n}
\eea
where $G$ is the Newton's constant with $\k^2\equiv 16 \pi G$; the Einstein tensor and the stress-energy tensor are defined respectively by
\bea
G_{\m\n}\equiv R_{\m\n}-\fr12 g_{\m\n}R
\eea
and
\bea
T_{\m\n}= \Big(F_{\m\r}F_\n{}^\r-\fr14 g_{\m\n}F^2 \Big)
\eea

\subsection{one-loop stress-energy tensor}

Although presenting a thorough analysis of the quantization procedure is not one of the present goals (because only the final outcomes will be needed for the analysis in the subsequent sections\footnote{Not all the steps of the quantization scheme of \cite{Park:2014tia} are needed because we are only interested in the one-loop analysis. For example, reduction of the physical states is not necessary to establish the one-loop renormalizability: the conventional method is sufficient in the presence of the cosmological constant \cite{Park:2016zgt}. Also, it is not obvious whether or not the quantization scheme could be applied to the three backgrounds considered in this work since the time-independent backgrounds are not, for example, asymptotically flat; more work is required to settle this matter.}), it will be useful to have a quantum-level perspective.
Before getting into the interwoven relationships among the quantization procedure, boundary conditions, loop effects and time-dependent solutions, we start with a brief account of the salient features of the quantization for the Einstein-Maxwell system. 
The content of this section is essential for the correct overall picture.

The quantization procedure brings to light a number of conceptual and technical issues not perceived in the past.
Let us start with the boundary terms and conditions. The issue of the boundary conditions has recently turned out to be much subtler than previously thought. The surface terms are important in several ways both at the classical and quantum levels. Here we focus on their quantization-related aspects, returning in what follows to the (better-known) subtleties in the definition of the classical stress-energy tensor. 
In conducting the action principle, one normally adds Gibbons-Hawking-type boundary terms by way of imposing the Dirichlet condition. 
It has recently been revealed that the Dirichlet boundary condition is just one of the possibly many boundary conditions to be collectively considered for the sake of a proper treatment of the entire Hilbert space.   
The status of this matter has several  implications.   
 One obvious implication is that it is now necessary to explore other types of boundary conditions. For instance, it was illustrated in \cite{Park:2015xoa} with the Einstein-Hilbert action that the boundary terms can be removed by the physical state conditions. Since this could be achieved without adding the Gibbons-Hawking term, the boundary condition was not restricted to the Dirichlet.  
Another not-so-obvious implication is that one must check whether or not the  classical-level boundary conditions are honored by the quantum corrections \cite{Park:2016fxc}. (In section 3 we will push further along this direction.)

A quite serious technical obstacle in the effective action computation is the complexity of the propagators associated with the curved backgrounds\footnote{To make matters worse, the effective action contains nonlocal terms in general. Such nonlocal terms could be important for the black hole physics at hand \cite{Mukhanov:1994ax}. They will not be considered in the present work for simplicity.}: they are known in closed forms only for a very limited number of cases. Thus it is difficult to conduct the perturbation theory around the actual curved background under consideration. Although not a stalemate, it makes it necessary to employ some additional measures such as covariance and dimensional analysis in order to determine the forms of the terms in the 1PI effective action. Also, since we are mostly interested in the ultraviolet divergences the flat space propagator can be employed to capture them.
One recent undertaking was the construction of the propagator out of the traceless components of the fluctuation metric \cite{Park:2015ota}. The necessity of employing the ``traceless" propagator is that the 4D covariance is maintained only when the traceless propagator is employed \cite{Park:2015ota}. An earlier related observation can be found, e.g., in \cite{Ortin}. 
The construction of the traceless propagator has been achieved in a manner convenient for the perturbative analysis. 
For a gravity-scalar system, the explicit one-loop analysis via employment of the traceless graviton propagator has been carried out in \cite{Park:2016zgt}. Similarly, the forms of the counter-terms in the case of the Einstein-Maxwell can be rather easily determined by a combination of  direct computation, dimensional analysis and 4D covariance.
With all these, one important aspect of the quantum effects is that the cosmological constant term is quite generically generated, regardless of the background under consideration \cite{Park:2015ota}.

At the quantum level, the stress-energy tensor computation should be done by starting with the renormalized action:
\bea
S=\int \sqrt{-g_r}\;\Big(\fr1{\k_r^2} R_r-\fr14 F_{r\m\n}^2 \Big)  \la{EM}
\eea
where the renormalized quantities are indicated by the subscript $r$.
After the one-loop analysis, the form of the 1PI effective action with the counter-terms takes (see \cite{Park:2015ota} for details\footnote{The counterterm computation of the Einstein-Maxwell action was of course done long ago. However, our recent finding shows that the correct determination of the coefficients requires use of the traceless propagator.}; earlier related analyses can be found, e.g., in \cite{Kallosh:1978wt,Capper:1984qq,Antoniadis:1995fc}) 
\bea
\hspace{-.1in}S=\fr1{\k^2}\int  \sqrt{-g}\Big[R-2\L\Big] -\fr14 \int \sqrt{-g}\;F_{\m\n}F^{\m\n}  
+\int  \sqrt{-g} \Big[ c_1 R^2+ c_2 R_{\m\n}R^{\m\n}+\cdots\Big] \nn\\ \la{qsactq} 
\eea
where $c$'s are constants - whose explicit values are not important for our purpose - that can be determined once the renormalization conditions are fixed.\footnote{Although the divergences can be determined by using a flat space propagator, the proper curved space propagator must be employed for the finite parts of the Feynman diagrams. The finite parts of the renormalized coefficients can then be fixed with a specific choice of a set of the renormalization conditions.} The cosmological constant has a purely quantum origin since the classical part was absent in \rf{EM}.
The part in the ellipsis contains the correction terms involving the Maxwell fields as well. In the presence of the cosmological constant, the  one-loop renormalizability can be established along the line of the conventional framework.

Above, the Riemann tensor square term, $R_{\m\n\r\s}R^{\m\n\r\s}$, appearing among the one-loop terms has been replaced by $R_{\m\n}R^{\m\n}$ and $R^2$ through the Euler-Gauss-Bonnet topological identity
\bea
R_{\m\n\r\s}R^{\m\n\r\s}-4R_{\m\n}R^{\m\n}+R^2=\mbox{total derivative} \la{Riemannsqid}
\eea 
As for the stress-energy tensor, there have been longstanding debates, even at the classical level, on its definition (see, e.g., \cite{Chen:1998aw,Kim:2013zha}). In general the surface terms matter for the stress-energy tensor, and they are responsible for part of the complications associated with its definition. A systematic treatment of the surface terms deserves work dedicated to itself and we will not attempt it here. One subtlety not as complicated is whether or not the one-loop-generated terms such as $R^2,R_{\m\n}^2$ should be included in the stress-energy tensor on the right-hand side of the metric field equation.
Considering the Bianchi identity associated with the Einstein tensor, it seems reasonable to place all of the quantum correction terms together with the matter part\footnote{This is also consistent with the definition of a stress-energy tensor given in \cite{Boulware:1985wk} in the context of the higher derivative gravity.}: the stress-energy tensor is obtained by taking the functional derivative of the matter part of the action with respect to the metric: 
\bea
\hspace{-6in} &&{ T_{\m\n}}=- \fr2{\k^2}\L g_{\m\n}+g_{\m\n}\Big[-\fr14 F^2   +\Big(c_1R^2-(4c_1+c_2)\nabla^2 R  +c_2 R_{\a\b}R^{\a\b}\Big) \Big]  \nn\\
&&\hspace{-.5in} +\Big[   F_{\m\r}F_\n{}^\r-2\Big(2c_1 RR_{\m\n}  -(2c_1+c_2) \nabla_\m \nabla_\n R
 -2c_2 R_{\k_1\m\n\k_2} R^{\k_1\k_2}+c_2\nabla^2 R_{\m\n}\Big)  +\cdots \Big]  \nn\\ \la{quanfeq}
\eea
In what follows we consider two backgrounds. The first is a Schwarzschild-Melvin solution \cite{Ernst}; the second is the recently found generalization of the Schwarzschild-Melvin solution \cite{Preston:2006ze}. Because the latter solution is more complex and less symmetric than the Schwarzschild-Melvin solution, it should provide a good test bed for one of the questions raised in the introduction.  
(The third case, to be considered in section 3, is an extension of the time-dependent AdS black hole analyzed at the classical level in \cite{Murata:2010dx} and at the quantum level in \cite{Park:2016fxc} and \cite{Park:2016vam}.)

\subsection{Schwarzschild-Melvin case}

The Schwarzschild-Melvin solution of the action
\bea
S=\int \sqrt{-g}\;\Big(\fr1{\k^2} R-\fr14 F_{\m\n}^2 \Big)
\eea
represent a Schwarzschild black hole immersed in an external magnetic field.
It is given by
\bea
ds^2=-f V^2 dt^2+\fr{V^2}{f}dr^2+r^2V^2d\th^2+\fr{r^2\sin^2\th}{V^2}d\f^2
\eea
where
\bea
f=1-\fr{2M}r \quad,\quad V=1+B^2r^2\sin^2\th;
\eea
the vector field is given by
\bea
A=\fr{B}{2{ \k}} \fr{1}{V(r,\th)} r^2\sin^2\th\; d\f   \la{vectorsol}
\eea
Since the solution represents a black hole inside a magnetic field of an infinite extent, it will be physical only when the combination $\fr{B}{\k}$ is small.
The coordinates $(t,\f)$ are cyclic and lead to the following first integrals: 
\bea
\dot{t}=\fr{E}{fG^2}\quad,\quad \dot{\f}=\fr{l V^2}{r^2\sin^2\th} \la{tfd}
\eea
where the dot means $\dot{}\equiv d/du$; $E,l$ are constants representing the conserved energy and angular momentum.
The geodesic $U^\m$ satisfies the normalization $U^\m U_\m=s$ where $s$ is $s=0,-1$ for null and timelike geodesics, respectively. The remaining second-order geodesic equations are presented in Appendix. The normalization can be written as
\bea
\frac{V^2 }{f} \dot{r}^2  -f V^2 \dot{t}^2+\frac{r^2 \sin ^2\theta }{V^2}  \dot{\phi }^2+ V^2 r^2 \dot{\theta }^2  =s   
\la{vsq} 
\eea
In principle, one should compute the geodesic up to (and including) the first subleading order in $\k^2$.
For the leading order, however, the quantum correction piece of the geodesic does not contribute when contracted with the stress-energy tensor, and one can therefore focus on the classical geodesic equations.

Let us consider the $\th=\pi/2$ case for which the equation above becomes substantially simplified. (The qualitative conclusion on the energy is not expected to change in more general cases.) With $\th=\pi/2$, eq.\rf{thdd2} is satisfied
and one can show
\bea
\dot{r}^2=\frac{s f}{V^2 \left(r,\frac{\pi }{2}\right)}+\frac{\text{E}^2}{V^4 \left(r,\frac{\pi }{2}\right)}-\frac{f l^2}{r^2} \la{rdsq}
\eea
We are now up to the task of computing the energy density as measured by a free-falling observer:
\bea
\r\equiv T_{\m\n}U_K^\m U_K^\n  \la{2dse}
\eea
where $U_K^\r$ denotes the four-velocity of an infalling observer in the Kruskal coordinates. $T_{\m\n} \equiv <K|\;T_{\m\n}^K \;|K>$ denotes the quantum-corrected stress tensor \rf{quanfeq} (reviews on the quantum-level stress tensor can be found in \cite{Candelas:1980zt,Birrell,Frolov,Mukhanov}): $T_{\m\n}^K$ represents the operator corresponding to the classical stress-energy tensor and $|K>$ denotes the Kruskal (i.e., Hartle-Hawking) vacuum.\footnote{The Schwarzschild vacuum (i.e., Boulware vacuum) was taken in \cite{Park:2014mba} for the Schwarzschild observer-oriented view. Here the energy computed by taking $|K>$ will represent the energy measured by the infalling observer.}

Let us examine the terms in \rf{quanfeq} to see whether or not they yield a high energy upon being contracted with the four velocities. Although the cosmological constant term comes with $\fr1{\k^2}$, its contribution to $\r$ should be small because of the small value of $\L$. Let us first consider the matter sector of the stress-tensor.
Upon evaluated at the classical background, the $F^2$ term in the stress-tensor yields
\bea
F^2=\fr{2 B^2}{\k^2}\frac{ (r-2 M)}{r \left(B^2 r^2+1\right)^4}
\eea
 and thus vanishes as $r\ra 2M$. As for the $F_{\m\r}F_\n{}^\r$ term, one can show that
 \bea
 F_{\m\r}F_\n{}^\r U^\m U^\n=\fr{B^2}{\k^2}\frac{ \left({E}^2 r-s \left(B^2 r^2+1\right)^2 (2 M-r)\right)}{r \left(B^2 r^2+1\right)^6}
 \eea 
thus
\bea
F_{\m\r}F_\n{}^\r U^\m U^\n  \ra  \fr{B^2}{\k^2} \frac{E^2  }{\left(4 B^2 M^2+1\right)^6}  \;\;\mbox{as} \;\; r\ra 2M
\eea 
Most of the gravity sector terms either identically vanish or vanish at the horizon. For example, one gets
\bea
 R=0 \quad , \quad R_{\m\n}R^{\m\n}= \frac{64 B^4 (-2M+r)^2}{r^2 \left(B^2 r^2 +1\right)^8}
 \quad , \quad R_{\k_1\m\n\k_2} R^{\k_1\k_2}=0
\eea
when evaluated at the classical background. As previously stated, the configuration is physical only when $\fr{B^2}{\k^2}$ is small and the energy encountered by an infalling observer will be moderate.

\subsection{generalized Schwarzschild-Melvin case}

A  new black hole solution with an asymptotically uniform magnetic field has been constructed in \cite{Preston:2006ze} by utilizing the so-called lightcone gauge. It is a two-parameter generalization of the Schwarzschild-Melvin solution and reduces to the Schwarzschild-Melvin spacetime in a certain parameter limit. Evidently it is more complex than the Schwarzschild-Melvin solution and should provide a test bed for examining the potential presence of the trans-Planckian energy.

The solution is obtained as a perturbation around the Schwarzschild black hole\footnote{We follow  \cite{Preston:2006ze} and use the Eddington-Finkelstein light-cone advanced coordinate $dv=dt+dr^*\equiv dt+dr/f(r)$. The exact form of $\tilde{g}_{\m\n}$ can be read off from eqs. (3.43)-(3.47) therein.} $ds^2=-f(r)dv^2+2 dv dr+r^2 d\W^2_\2$ with $f(r)=1-\fr{2M}{r}$:
\bea
&& g_{vv}=-f(r)+\gt_{vv}(r,\th),\quad  g_{vr}=1, \quad g_{v\th}=\gt_{v\th}(r,\th) \nn\\
&& \;\;g_{\th\th}=r^2+\gt_{\th\th}(r,\th),\quad g_{\vf\vf}=r^2 \sin^2\th+\gt_{\vf\vf}(r,\th)
\la{gvfvfpert}
\eea
Just as in the Schwarzschild-Melvin case, it is useful to find the first integrals. For this, note that the coordinates $t,\f$ are again cyclic, leading to the following first integrals:
\be
E=-g_{vv} \dot{v}-g_{vr} \dot{r}-g_{v\th}\dot{\th} ,
\la{E}
\ee
\be
l=g_{\vf\vf} \dot{\vf}.
\la{l0}
\ee
where again $\dot{}\equiv d/du$ and $E,l$ are constants. The remaining geodesic equations can be found in Appendix A. The geodesic normalization condition reads, to the first order in the perturbed metric,
\be
g_{\m\n} U^\m U^\n=(-f(r)+\tilde{g}_{vv})\dot{v}^2+2\dot{v}\dot{r}+2\dot{v}\dot{\th}\tilde{g}_{v\th}+(r^2+\tilde{g}_{\th\th})\dot{\th}^2+(r^2\sin^2\th+\tilde{g}_{\vf\vf})\dot{\vf}^2=s
\la{v^2def}
\ee
Things get quite simplified by choosing $\th=\pi/2$; after some algebra (more details in Appendix) one gets
\bea
\dot{r}^2 
&=& E^2+s(f-\tilde{g}_{vv})-\fr{l_0^2}{r^2} (f-\tilde{g}_{vv}-\fr{f}{r^2} \tilde{g}_{\vf \vf})+{\cal O}(\tilde{g}^2)
\la{dotr^2}
\eea
With the help of the Mathematica package diffgeo.m, it is checked that the computation of $\r$ does not yield a trans-Planckian energy in this case. For example, one gets, for the $F_{\m\r}F_\n{}^\r $ term, 
\bea
F_{\m\r}F_\n{}^\r U^\m U^\n  \ra { E^2} \fr{B^2}{\k^2} \;\;\mbox{as} \;\; r\ra 2M
\eea 
As before the presence of the small parameter $\fr{B^2}{\k^2}$ makes this contribution small.

\section{Trans-Planckian energy}

Although a more systematic and complete study of boundary conditions is still to be carried out in gravity quantization, it is nevertheless possible to probe the role of the boundary modes in the dynamical evolution of the system. In this section we deepen our understanding of the case whose analysis has been carried out to some extent in \cite{Park:2016fxc} and \cite{Park:2016vam}; it was shown that the quantum gravitational effects and non-Dirichlet modes (to be defined) lead to a time-dependent solution. 
After reviewing \cite{Park:2016fxc} and  \cite{Park:2016vam} in section 3.1, we extend the analysis by focusing, for one thing, on the loop-corrected cosmological constant. The trans-Planckian energy does not arise at the classical level. This very fact may not be so surprising. However, the detailed manner in which this happens is surprising. We show that the quantum-level solution does display a trans-Planckian energy.

The classical action we consider in this section is
\bea
 S=\fr1{\k^2}\int d^4 x \sqrt{-g}\Big[R-2\L\Big] 
-\int d^4 x \sqrt{-g}\Big[\fr12(\pa_\mu \z)^2 +\fr12m^2\z^2\Big] \la{csa}
\eea
It admits an AdS black hole solution, 
\bea
\z=0\quad,\quad ds^2=-\fr1{z^2}\Big(Fdt^2+2dtdz \Big)+\Phi^2(dx^2+dy^2)   \la{cs}
\eea
with
\bea
F=-\fr{\L}{3}-2M z^3\quad,\quad \Phi=\fr1{z} \quad,\quad \z=0
\eea

\subsection{time-dependent solution of gravity-scalar system}

A gravity-scalar system was considered at the quantum level in \cite{Park:2016fxc}  and \cite{Park:2016vam}.
The one-loop 1PI effective action after one-loop renormalization of  the classical action \rf{csa} is  
\bea
&&\hspace{.1in} S=\fr1{\k^2}\int d^4 x \sqrt{-g}\Big[R-2\L\Big] 
-\int d^4 x \sqrt{-g}\Big[\fr12(\pa_\mu \z)^2 +\fr12m^2\z^2\Big]\nn\\
&&\hspace{-.3in}+\fr1{\k^2}\int d^4 x \sqrt{-g}\Big[e_1{ \k^4} R\z^2+e_2 \k^2R^2+e_3\k^2 R_{\m\n}R^{\m\n}+e_4 { \k^6}(\pa\z)^4+e_5{ \k^6} \z^4+\cdots\Big]  \la{qsactscalar} \nn\\
\eea
where $e$'s are constants that can be fixed with fixed renormalization conditions.
The metric and scalar field equations that follow from \rf{qsactscalar} are
\bea
&& R_{\m\n}-\fr12Rg_{\m\n}+\L g_{\m\n}-\fr12g_{\m\n}\Big(-\fr12 { \k^2}(\pa_\mu \z)^2 -\fr12m^2 { \k^2}\z^2   \nn\\
&&\hspace{-.5in}+e_1 { \k^4}R\z^2 -2 e_1{ \k^4}\nabla^2\z^2 +e_2{ \k^2}R^2-(4e_2+e_3){ \k^2}\nabla^2 R  +e_3{ \k^2} R_{\a\b}R^{\a\b} +e_4{ \k^6} (\pa\z)^4+e_5{ \k^6} \z^4\Big)  \nn\\
&&\hspace{-.1in}-\fr12 { \k^2}\pa_\m \z \pa_\n\z+e_1{ \k^4}R_{\m\n}\z^2-e_1 { \k^4}\nabla_\m\nabla_\n \z^2+2e_2{ \k^2} RR_{\m\n}  -(2e_2+e_3){ \k^2} \nabla_\m \nabla_\n R\nn\\
&& \hspace{-.1in} -2e_3{ \k^2} R_{\k_1\m\n\k_2} R^{\k_1\k_2}+e_3{ \k^2}\nabla^2 R_{\m\n}  +2e_4 { \k^6}\pa_\m \z \pa_\n\z (\pa\z)^2=0   \la{quanfe}
\eea
\bea
\nabla^2\z-m^2\z+2e_1\k^2 R\z-4e_4\k^4\Big[\nabla^2\z \, (\pa\z)^2+2\nabla_\a\z\,(\nabla^\a\nabla^\b\z)\,\nabla_\b\z\Big]+4e_5\k^4 \z^3=0\nn
\eea
The field equations above can be solved by employing the metric ansatz \cite{Murata:2010dx}
\bea
ds^2=-\fr1{z^2}\Big(F(t,z)dt^2+2dtdz \Big)+\Phi^2(t,z)(dx^2+dy^2)  \la{ma}
\eea
with the following quantum-corrected series
\bea
\hspace{-.5in}F(t,z)&=& F_0(t) +F_1(t) z+ F_2(t)z^2+F_3(t)z^3 + ...\nonumber\\
&+&\k^2 \Big[F_0^h(t) +F_1^h(t) z+ F_2^h(t)z^2+F_3^h(t)z^3 + ...\Big] \nn\\
\Phi(t,z)&=&\frac{1}{z}+\Phi_0(t) +\Phi_1(t) z+ \Phi_2(t)z^2+\Phi_3(t)z^3 + ...\nonumber\\
&+& \k^2\Big[\fr{\Phi_{-1}^h(t)}{z}+\Phi_0^h(t) +\Phi_1^h(t) z+ \Phi_2^h(t)z^2+\Phi_3^h(t)z^3 + ...\Big] 
   \la{1stans}
\eea
where the modes with superscript `$h$' represent the quantum modes. The quantum corrections of the metric imply a deformation of the geometry by quantum effects \cite{Park:2013rm,Park:2017dib}. (See also \cite{Giddings:2017mym}\cite{Dey:2017yez}\cite{Guo:2017jmi} for related works.) Similarly, for the scalar field,
\bea
\z(t,z)&=&\z_0(t) +\z_1(t) z+ \z_2(t)z^2+\z_3(t)z^3 + ...\nonumber\\
&+&\k^2\Big[ \z_0^h(t) +\z_1^h(t) z+ \z_2^h(t)z^2+\z_3^h(t)z^3 + ...\Big] \la{zetaser}
\eea
It was found that the Dirichlet boundary condition is not preserved by the quantum corrections. Different boundary conditions can be adopted by adjusting the boundary modes. For example, one imposes $\Phi_0(t)= 0, \Phi_0^h(t)= 0$ and $\Phi_{-1}^h(t)=0$ for the Dirichlet boundary condition. The modes such as $\Phi_0(t),\Phi_0^h(t),\Phi_{-1}^h(t)$ will be called the ``non-Dirichlet modes" for this reason. The following choice - which corresponds to a non-Dirichlet boundary condition - was explored: 
\bea
\Phi_0(t)\neq 0\quad,\quad \Phi_0^h(t)\neq 0 \la{nDiricon}
\eea
By analyzing the field equations expanded in the $z$-series, one can show,
for the classical modes,
\bea
&&m^2=\fr{2\L}{3},\quad \z_0=0,\quad { F_0=-\fr{\L}{3}},\quad  \Phi _1=0,\quad { F_1=-F_0 \Phi _0-\Lambda \Phi _0} \nn\\
&&\Phi _2=0,\quad F_2=\frac{1}{4} \left(4 F_0 \Phi _0{}^2-8 \dot{\Phi} _0 \right)\nn\\
&&\z_3=0,\quad \Phi_3=0,\quad {F}_3=const,\quad  \z_4=0,\quad \Phi_4=0  ,\quad F_4=-F_3\Phi_0;  \nn\\
 \la{mrone}
\eea
for quantum modes 
\bea
&&\hspace{-.3in}\z_0^h=0,\quad  F_0^h=0,\quad
\Phi_{1}^h=0,\quad { F_1^h}=\frac{2}{3} \Big(3 {F}_0^h \Phi _0+{\Lambda} \Phi _0 {\Phi}_{-1}^h-{\Lambda} {\Phi}_0^h-3 \dot{\Phi}_{-1}^h \Big), \nn\\
&&\hspace{.3in}\Phi_2^h=0, \quad F_2^h=  \frac{1}{3} \Big(2 {\Lambda} \Phi _0{}^2 {\Phi}_{-1}^h -2 {\Lambda} \Phi _0 {\Phi }_0^h+6 {\Phi}_{-1}^h \dot{\Phi} _0-6 \dot{\Phi}_0^h  \Big),
\nn\\
&&\hspace{-.5in} \Phi_3^h=0,    \quad { \dot{F}_3^h}=-3F_3\dot{\Phi}_{-1}^{h}, \quad  { \z_3^h}=   -\fr1{\text{$\Lambda$}}  \Big({\text{$\Lambda $} \text{$\zeta $}_1^h \Phi _0{}^2+ 2 \text{$\Lambda $} \text{$\zeta $}_2^h \Phi _0+ 3 \Phi _0 \text{$\dot{\zeta} $}_1^h+ 3 {\zeta}_1^h \dot{\Phi} _0+  3 \text{$\dot{\zeta} $}_2^h} \Big)    
\nn\\
&&\hspace{-.3in} F_4^h=F_3 \Phi _0 \Phi_{-1}^h-F_3 \Phi_0^h-F_3^h \Phi _0,\quad
\Phi_4^h=-3 {e_2} F_3 \Phi _0{}^2+3 {e_2} F_5-2 {e_3} F_3 \Phi _0{}^2+2 {e_3} F_5\nn\\
&&\hspace{-.5in}\z_4^h=  \frac{F_3 \zeta_1^h}{2 \text{$\Lambda $}} +\frac{12 \dot{\zeta}_1^h \dot{\Phi} _0}{\Lambda^2}+\frac{6 \Phi _0 \ddot{\zeta}_1^h}{\Lambda^2}+\frac{6 \zeta_1^h \ddot{\Phi} _0}{\Lambda^2}+\frac{6 \ddot{\zeta}_2^h}{\Lambda^2}
+\frac{9 \Phi _0{}^2 \dot{\zeta}_1^h}{\text{$\Lambda $}}+\frac{9 \Phi _0 \dot{\zeta}_2^h}{\text{$\Lambda $}}+\frac{9 \zeta_1^h \Phi _0 \dot{\Phi} _0}{\text{$\Lambda $}}+2 \zeta_1^h \Phi _0^3+3 \zeta_2^h \Phi _0^2  \nn\\ \la{mrtwo}
\eea
An intriguing finding was that the quantum-level analysis actually imposes additional constraints on the classical modes. (The field equations have the terms of order $\hbar$ since all of the coefficients $e$'s in \rf{quanfe} come with $\hbar$, and once the series ansatze \rf{1stans} and \rf{zetaser} are substituted, the classical modes such as $\z_1,\z_2$ come to appear in the parts of the equation of $\hbar$ order, which leads to additional constraints among the classical modes. We will come back to this in the conclusion.) Since this is an important point we elaborate: according to the classical analysis \cite{Murata:2010dx}, the modes $\z_1,\z_2$ are free and responsible for the entire dynamics as the higher modes are given in terms of $\z_1,\z_2$ and their derivatives. However, the quantum-level analysis unravels that the two modes become constrained:
\bea
\z_1=0=\z_2 \la{z1z20}
\eea
On the contrary, the quantum-counterpart modes, $ \z_1^h$ and $ \z_2^h$, are not constrained. As a matter of fact, with $\Phi_0(t), \Phi_0^h(t)$ and $\Phi_{-1}^h(t)$ they determine the higher modes; namely, the higher modes become functions of these modes. 

Let us pause and ponder the implications of the results.
Firstly, the solution represents the quantum-modified time-dependent black hole solution, and the quantum modes above are the ones that feed the time-dependence of the solution.
Secondly, the presence of such modes implies that the quantum-corrected solution no longer satisfies the Dirichlet condition. Their presence also implies nontrivial dynamics on the boundary where part of the system information is stored. 
The third implication is perhaps even more intriguing. The time-dependence of the classical black hole solution with a Dirichlet boundary condition is an apparent phenomenon: were it not for the presence of the quantum modes, the quantum-level constraints force the solution to reduce to a {\em time-independent} configuration, namely, an AdS black hole when the classical non-Dirichlet mode $\Phi_0$ is absent.

Before we proceed, let us note a curious resemblance to the finding in \cite{Park:2017dib} where the time-dependent solution constructed in \cite{Chadburn:2013mta} was checked against a trans-Planckian energy. There, elimination of the cosmological constant term made the time-dependence disappear. In the case of \cite{Park:2016fxc} and \cite{Park:2016vam} just reviewed, what feeds the time-dependent solution is the non-Dirichlet modes. As we will soon see, it is not only the non-Dirichlet modes but also the quantum-corrected cosmological constant that feeds the time-dependence which in turn will be crucial for the trans-Planckian energy.

\subsection{extension by quantum cosmological constant}

The analysis in \cite{Park:2016fxc} and \cite{Park:2016vam} did not take the quantum corrections of the cosmological constant. In other words, the cosmological constant $\L$ was taken to be entirely classical. Here we extend the analysis by focusing on the effects of the loop-corrected cosmological constant, writing it explicitly as $\L\equiv \L_0+\k^2\L_1$ with $\L_0,\L_1$ classical and quantum, respectively. With this, slightly modified mode relations are obtained; although the modifications are modest, the implications are not insignificant and several interesting aspects of the dynamics are revealed. For instance, the quantum-induced cosmological constant contributes to the time dependence of the solution. 

The procedure of solving the field equations goes the same apart from having to include the quantum correction piece of the cosmological constant. For the classical modes, one gets
\bea
&&m^2=\fr{2\L_0}{3},\quad \z_0=0 ,\quad { F_0=-\fr{\L_0}{3}},\quad \Phi _1=0,\quad {F_1=-F_0 \Phi _0-\Lambda_0 \Phi _0} \nn\\
&&\hspace{.5in}\Phi _2=0,\quad F_2=\frac{1}{4} \left(4 F_0 \Phi _0{}^2-8 \dot{\Phi} _0 \right)\nn\\
&& \z_3=0,\quad \Phi_3=0  ,\quad {F}_3=const,\quad   \z_4=0,\quad \Phi_4=0 ,\quad  F_4=-F_3\Phi_0 ;  \nn\\
 \la{mrone2}
\eea
for the quantum modes, 
\bea
&&\hspace{-.3in}\z_0^h=0,\quad { F_0^h=-\frac{1}{3} \kappa ^2 \Lambda_1} ,\quad
\Phi_{1}^h=0,\quad { F_1^h}=\frac{2}{3} \Big(3 {F}_0^h \Phi _0+{\Lambda_0} \Phi _0 {\Phi}_{-1}^h-{\Lambda_0} {\Phi}_0^h-3 \dot{\Phi}_{-1}^h \Big), \nn\\
&&\Phi_2^h=0, \quad { F_2^h}=  \frac{1}{3} \Big(-\kappa ^2 {\Lambda_1} \Phi _0{}^2+2 {\Lambda_0} \Phi _0{}^2 {\Phi}_{-1}^h -2 {\Lambda_0} \Phi _0 {\Phi }_0^h+6 {\Phi}_{-1}^h \dot{\Phi} _0-6 \dot{\Phi}_0^h  \Big),
\nn\\
&&\hspace{-.5in}\quad  { \z_3^h}=   -\fr1{\text{$\Lambda_0$}}  \Big({\text{$\Lambda_0 $} \text{$\zeta $}_1^h \Phi _0{}^2+ 2 \text{$\Lambda_0 $} \text{$\zeta $}_2^h \Phi _0+ 3 \Phi _0 \text{$\dot{\zeta} $}_1^h+ 3 \text{$\zeta $}_1^h \dot{\Phi} _0+  3 \text{$\dot{\zeta} $}_2^h} \Big)    ,     \quad   \Phi_3^h=0,\quad  { \dot{F}_3^h}=-3F_3\dot{\Phi}_{-1}^{h}
\nn\\
&&\hspace{-.3in} F_4^h=F_3 \Phi _0 \Phi_{-1}^h-F_3 \Phi_0^h-F_3^h \Phi _0,\quad
\Phi_4^h=-3 {e_2} F_3 \Phi _0{}^2+3 {e_2} F_5-2 {e_3} F_3 \Phi _0{}^2+2 {e_3} F_5\nn\\
&&\hspace{-.5in} \z_4^h=  \frac{F_3 \zeta_1^h}{2 \text{$\Lambda_0 $}} +\frac{12 \dot{\zeta}_1^h \dot{\Phi} _0}{\Lambda_0^2}+\frac{6 \Phi _0 \ddot{\zeta}_1^h}{\Lambda_0^2}+\frac{6 \zeta_1^h \ddot{\Phi} _0}{\Lambda_0^2}+\frac{6 \ddot{\zeta}_2^h}{\Lambda_0^2}+\frac{9 \Phi _0{}^2 \dot{\zeta}_1^h}{\text{$\Lambda_0 $}}+\frac{9 \Phi _0 \dot{\zeta}_2^h}{\text{$\Lambda_0 $}}+\frac{9 \zeta_1^h \Phi _0 \dot{\Phi} _0}{\text{$\Lambda_0 $}} \nn\\ 
&&\hspace{3in}+2 \zeta_1^h \Phi _0^3+3 \zeta_2^h \Phi _0^2
 \la{mrtwo2}
\eea
Several salient features of the outcome are as follows. The result above shows that in order for, e.g., $F_2$ not to vanish, the presence of the non-Dirichlet mode $\Phi _0(t)$ is important. To see things more clearly, let us set the entire quantum modes aside. As can be seen from \rf{mrone2} there still exists a time-dependent solution (that can be consistently extended to the quantum level) if one keeps the non-Dirichlet mode $\Phi_0(t)$. The Dirichlet condition tends to suppress the time dependence: suppose the quantum mode $\Phi _{-1}^h(t)$ and its derivative $\dot{\Phi} _{-1}^{h}(t)$ are absent. Then $F_1^h(t)$ would vanish if $ \Phi _0(t)$ is absent as well. This shows that the non-Dirichlet modes and the quantum corrections together feed the time-dependence of the solution. (More on the non-Dirichlet modes in the conclusion.) Also, $\L_1$ contributes to $F_2^h$; this shows that the quantum-induced cosmological constant too contributes to the time dependence of the solution.

The following will be important for the energy analysis in the next subsection. As stated in the previous subsection the classical time-dependent solution of \cite{Murata:2010dx} is `demoted' to an AdS black hole by the quantum-level constraints. The classical-level time-dependence of the solution of \cite{Murata:2010dx} is not preserved at the quantum level: the quantum-level constraints force the classical part of the resulting solution to become an AdS black hole that is time-independent at the classical level. Put differently, additional constraints among the classical modes arise at the quantum level. Once those constraints are enforced on the classical part of the solution, the classical metric becomes that of the usual time-independent AdS black hole.

\subsection{trans-Planckian energy}

Let us compute the energy density measured by a free-falling observer, $\r\equiv T_{\m\n}U^\m U^\n$,
where $T_{\m\n}$ denotes the quantum-corrected stress tensor \rf{quanfeq} and $U^\m$ the geodesic. 
As in section 2, we first work out the geodesics. The geodesics for the {\em classical} AdS black hole can be used for the purpose of computing $\r$ for the reason which will become clearer below. The stress-energy tensor must be evaluated at the quantum-corrected solution.
Since we are ultimately interested in the energy near the horizon, we will, at some point, consider the solution in the $z-z_{EH}$ series where $z_{EH}$ denotes the location of the classical horizon. 

The classical part of the solution obtained in the previous subsection is time-dependent in general due to the presence of the non-Dirichlet mode $\Phi_0(t)$, and this causes unnecessary complications in finding the geodesic. We thus choose
\bea
F_3=-2M\;\; \mbox{and set}\;\; \Phi_0(t)=0
\eea 
 Since $\z_1=0=\z_2$, the classical part of the full quantum-level solution is the same as the well-known one given in \rf{cs}. (Nevertheless, the overall solution will be a time-dependent one due to the presence of the time-dependent quantum modes.) Although this is a special case, it is expected to share the important features of a more general solution when it comes to the trans-Planckian scaling of the energy. With this, the classical geodesic can be computed straightforwardly.
From the metric of the AdS black hole\footnote{We have set $\L=-3$ by following the common practice in the literature.}
\bea
ds^2=-\fr1{z^2}\Big[(1-2M z^3)dt^2 +2dtdz\Big]+\fr1{z^2}(dx^2+dy^2)
\eea
where $M$ is a parameter proportional to the mass of the black hole, the first integrals follow:
\bea
\fr{1}{z^2} \Big[(1-2Mz^3)\dot{t} +\dot{z}\Big]=E \quad,\quad  \fr1{z^2}\dot{x}=l_1 \quad,\quad \fr1{z^2}\dot{y}=l_2
\eea
With these, the velocity normalization condition, $U^2=s$, takes
\bea
\dot{z}^2=s z^2+(E^2-l_1^2-l_2^2)z^4-2sMz^5+2M(l_1^2+l_2^2)z^7
\eea 
and one gets 
\bea
\dot{t}=\fr1{1-2Mz^3}\Big(Ez^2+\sqrt{s z^2+(E^2-l_1^2-l_2^2)z^4-2s Mz^5+2M(l_1^2+l_2^2)z^7}\Big)\nn\\
\eea
The one-loop stress-energy tensor is given by
\bea
&&T_{\m\n} =  \pa_\m \z \pa_\n\z   -g_{\m\n}\Big({ \fr1{\k^2}}\Lambda +\fr12 (\pa_\mu \z)^2 + \fr12 m^2 \z^2 \Big) \nn\\  
&&\hspace{-.5in}+\k^2e_1 (-2 R_{\m\n}\z^2+2 \nabla_\m\nabla_\n \z^2) +e_1\k^2 g_{\m\n}(
R\z^2 -2 \nabla^2\z^2  ) +\cdots  \la{Tquanq}
\eea
Let us focus on the leading order terms in the first line. (As before we disregard the cosmological constant term in the stress tensor.) The second term in the first line is bound by the geodesic normalization, $U_\m U^\m=s$, thus of subleading order. Given the structure of $\dot{t}$ above, the first term, namely the scalar kinetic term $\pa_\m \z \pa_\n\z$ can potentially yield a large value of the energy. In other words, at least naively, a large value of the energy is expected to come from the $\dot{t}$ components of $\r$ since $\dot{t}$ scales as $\dot{t} \sim \fr1{1-2Mz^3}$ and the classical horizon $z_{EH}$ is located at the vanishing of $1-2Mz^3$,
$z_{EH}^3\equiv \fr1{2M}$.  
It is possible at this point to see why a classical, as opposed to one-loop, geodesic is sufficient for our purpose, a statement made earlier. Let us examine the contribution of the first term in \rf{Tquanq} to $\r$, $\pa_\m \z \pa_\n\z \;U^\m U^\n$. With $\dot{t}\sim \fr{1}{z-z_{EH}}$ it is the $\dot{\z}\dot{\z}\dot{t} \dot{t}$ piece that will give the leading order energy. From this it follows that the classical part of the geodesic is sufficient for obtaining one-loop $\r$: because the quantum-level field equations constrain $\z_1,\z_2$ such that $\z_1=0=\z_2$, the time-dependent part of the solution for the field equations is only the quantum correction piece. Since the stress-energy tensor part - namely $\pa_\m \z \pa_\n\z$ - is already second order in $\hbar$ (that we have been suppressing) and $\k^2$, the geodesic for the classical AdS black hole is sufficient.

For the remainder of this subsection, we examine the $\k$-scalings of various quantities to determine the scaling of the energy. 
At least to the orders analyzed in \cite{Park:2016vam} and reviewed above,  the classical piece of the scalar field is absent: the original scalar field expansion 
\bea
\z(t,z)&=& \z_0(t)+ \z_1(t) z+ \z_2(t)z^2+\z_3(t) z^3 +\cdots  \nn\\
  && +\k^2\Big[\z_0^h(t)+\z_1^h(t) z+ \z_2^h(t)z^2+\z_3^h(t) z^3 +\cdots \Big]
\la{zsero}
\eea
reduces, on account of \rf{mrone} and \rf{mrtwo}, to
\bea
\z(t,z)&=& \k^2\Big[ \z_1^h(t) z+ \z_2^h(t)z^2+\z_3^h(t) z^3 +\cdots  \Big] \la{zetaser2}
\eea
where the modes $\z_1^h(t),  \z_2^h(t)$ are free (i.e., unconstrained) and the expression for, e.g., $\z_3^h(t)$ can be found in \rf{mrtwo2}. The vanishing of the classical piece will bear important implications for the energy so we run a double-check to ensure that it remains true to all orders in $z$, not just to the first several orders explicitly checked. To this end and for a more transparent understanding of the behavior of the scalar near the horizon, let us re-expand the $z$-series solution in $z-z_{EH}$.
The re-expansion of \rf{zetaser2} around $z_{EH}$ takes, a priori, 
\bea
\z(t,z)&=& \tilde{\z}_0(t) +  \tilde{\z}_1(t) (z-z_{EH})+ \tilde{\z}_2(t)(z-z_{EH})^2+\tilde{\z}_3 (z-z_{EH})^3 +\cdots\nn\\
&&\hspace{-.5in}+\k^2\Big[\tilde{\z}_0^h(t) +  \tilde{\z}_1^h(t) (z-z_{EH})+ \tilde{\z}_2^h(t)(z-z_{EH})^2+\tilde{\z}_3^h (z-z_{EH})^3 +\cdots \Big]  \nn\\
\la{Zser}
\eea
Given $\dot{t}\sim \fr{1}{z-z_{EH}}$, a potentially large value of the energy will arise from the term of the $(z-z_{EH})^0$ order, $\tilde{\z}_0(t)$. As for the quantum mode $\tilde{\z}_0^h(t)$, it comes with a $\k^2$ factor and is set aside for now (we will come back to it below); let us focus on the classical mode $\tilde{\z}_0(t)$. Since \rf{Zser} is a re-expansion of \rf{zsero}, $\tilde{\z}_0(t)$ will be given by sum of the original modes $\z_n$'s with $n\geq 0$. 
By running the program that led to \rf{mrone2} and \rf{mrtwo2} but now in the new series, one can show that $\tilde{\z}_0(t)=0$.\footnote{ 
For this, it is convenient to introduce
\bea
Z\equiv z-z_{EH}
\eea
and rewrite \rf{ma} as
\bea
ds^2=-\fr1{(Z+z_{EH})^2}\Big(\Ft(t,Z)dt^2+2dtdZ \Big)+\tilde{\Phi}^2(t,Z)(dx^2+dy^2)  \la{maZ}
\eea
with
\bea
\hspace{-.5in}\Ft(t,z)&=& \Ft_0(t) +\Ft_1(t) Z+ \Ft_2(t)Z^2+\Ft_3(t)Z^3 + ...\nonumber\\
&& +\k^2 \Big[\Ft_0^h(t) +\Ft_1^h(t) Z+ \Ft_2^h(t)Z^2+\Ft_3^h(t)Z^3 + ...\Big] \nn\\
\Phit(t,z)&=&\Phit_0(t) +\Phit_1(t) Z+ \Phi_2(t)Z^2+\Phit_3(t)Z^3 + ...\nonumber\\
&&+ \k^2\Big[\Phit_0^h(t) +\Phit_1^h(t) Z+ \Phit_2^h(t)Z^2+\Phit_3^h(t)Z^3 + ...\Big] \nn\\
 \zetat(t,z)&=&\zetat_0(t) +\zetat_1(t) Z+ \zetat_2(t)Z^2+\zetat_3(t)Z^3 + ...\nonumber\\
&&+\k^2\Big[ \zetat_0^h(t) +\zetat_1^h(t) Z+ \zetat_2^h(t)Z^2+\zetat_3^h(t)Z^3 + ...\Big] \la{zetasert}
\eea
}
The fact that $\tilde{\z}_0(t)$ vanishes implies that the vanishing of the classical part of the scalar solution, although established to the first several orders in the original $z$-series, remains valid to all orders. More specifically, the finding that the higher modes ${\z}_n$ with $n\geq 3$ are functions of $\z_1, \z_2$ must remain valid to all orders, and thus all of the higher modes $\z_n$ vanish.

The fact that the matter part of the action comes at higher order of $\k^2$ translates into the form of the metric field equation where the matter part starts with at $\k^2$ order:
\bea
 R_{\m\n}-\fr12Rg_{\m\n}+\L g_{\m\n}+\fr{\k^2}2g_{\m\n}\Big(\fr12 (\pa_\mu \z)^2 +\fr12m^2 \z^2  \Big) -\fr12 { \k^2}\pa_\m \z \pa_\n\z+\cdots=0   \nn\\
\eea
This implies that the solution generically takes a form of
\bea
\z=\fr{\xi}{\k}
\eea
where $\xi$ represents the rescaled scalar field. Since the classical part identically vanish, 
$\xi(t,z)$ has the following series:
\bea
\xi(t,z)={\k^2}\Big[
\tilde{\xi}_0^h(t) +  \tilde{\xi}_1^h(t) (z-z_{EH})+ \tilde{\xi}_2^h(t)(z-z_{EH})^2+\tilde{\xi}_3^h (z-z_{EH})^3 +\cdots \Big]\nn\\
\la{Zsert1}
\eea
which implies
\bea
\z(t,z)={\k}\Big[
\tilde{\xi}_0^h(t) +  \tilde{\xi}_1^h(t) (z-z_{EH})+ \tilde{\xi}_2^h(t)(z-z_{EH})^2+\tilde{\xi}_3^h (z-z_{EH})^3 +\cdots \Big]\nn\\
\la{Zsert2}
\eea
Let us consider the scalar kinetic term in the stress-energy tensor and its contribution to $\r$, $\pa_\m \z \pa_\n\z \;U^\m U^\n$. At the classical level, $\dot{t}$ scales as
\bea
\dot{t}=-\fr{E}{3M} \fr{1 }{z-z_{EH}}+{\cal O}((z-z_{EH})^0)
\eea
The location of the horizon at the quantum level, $z_{EH}^q$, will take a form of
\bea
z_{EH}^q=z_{EH}+ {\cal O}(\k^{2})
\eea 
and this implies
\bea
\dot{t}\sim {\cal O}(\k^{-2})
\eea
at $z=z_{EH}^q$. With this scaling it is the $\dot{\z}\dot{\z}\dot{t} \dot{t}$ piece of $\r$ that will give the leading order energy. As $z\ra z_{EH}^q$, one gets
\bea
\pa_\m \z \pa_\n\z \;U^\m U^\n \sim \fr{\k^2 [\dot{\tilde{\xi}}_0^h(t)]^2}{\k^4} \sim \fr1{\k^2}
\eea
Note that it is the ``horizon quantum mode" $\tilde{\xi}_0^h(t)$ that led to this trans-Planckian energy. What appears above is a time derivative of $\tilde{\xi}_0^h(t)$; a time-independent mode $\tilde{\xi}_0^h(t)=const$ will not lead to a trans-Planckian energy.
The boundary modes are the important part of the physical degrees of freedom and must hold part of the system information. They determine the bulk dynamics as analyzed in the previous subsections. More basically, they are the building blocks of the time-dependence and represent the boundary dynamics and deformations. The result above shows that being a part of the horizon mode, they are also linked with the trans-Planckian energy.

\section{Conclusion}

In this sequel, we have further explored the intertwined relationships among boundary conditions, quantum effects, and time-dependent solutions. Three black hole backgrounds have been analyzed: a Schwarzschild-Melvin black hole, its generalization obtained in \cite{Preston:2006ze} and the generalization of the time-dependent AdS black hole considered in \cite{Murata:2010dx} and \cite{Park:2016vam}. A pattern similar to that of \cite{Park:2017dib} has again been found: the non-Dirichlet modes and quantum effects are crucial for a quantum-modified time-dependent black hole solution.
One of our main focuses is on the quantum-induced cosmological constant and it is shown that it is one of the agents that reinforce the time-dependence of the solution.\footnote{The importance of the cosmological constant was discussed in a different context in \cite{Ho:2017vgi} in which the role of the back reaction of the vacuum energy in the black hole geometry was noted.} The trans-Planckian energy is obtained in the case of the time-dependent solution.

It is confirmed that the time-dependence of the solution is at odds with the Dirichlet boundary conditon. 
The boundary conditions are closely tied with quantization procedure. It is rather surprising that adoption of  such an innocuous boundary condition as the Dirichlet leads to a (presumably highly) limited subset of the proper Hilbert space. This phenomenon is in no way elementary: one does not have an analogous phenomenon with a system where the metric is kept as non-dynamical.  The limitation of the Dirichlet boundary condition has its origin in the fact that the physical degrees of freedom of a gravitational theory happens to be associated with the hypersurface at the boundary, and thus get suppressed by the Dirichlet boundary condition.

We believe that the present work with the previous ones unequivocally shows that the quantum gravitational effects cannot in general be disregarded, especially in time-dependent circumstances, since they can be important for nonpertubative physics. It has been shown that with the quantum-level constraints taken into account, the classical time-dependent black hole solution ``reduces" to the AdS black hole solution in the sense explained in the main body. Also, it is  the quantum gravitational effects that lead to the trans-Planckian energy as demonstrated in the main body.\footnote{It will be interesting to see whether the quantum-induced trans-Planckian energy is responsible for extreme high energy gamma rays from active galactic nuclei.}

The phenomenon seen in \rf{z1z20} seems to have its origin in the subtlety of going to classical limit \cite{Ballentine}. In the present case, the subtley is manifest as follows. As the $\hbar$-order parts of the field equations must vanish separately from the classical parts, one gets
\bea
\hbar(\cdots)=0
\eea
Inside the parenthesis, some of the classical modes come to appear. If one takes $\hbar\ra 0$-limit too early, the quantum-level constraint will be removed and this corresponds to the ``usual" classical limit. As our analysis explicitly shows, the full quantum-level analysis can (and in our case, it does) introduce ``order 1" changes to the classical solution through the constraints coming from the part represented by the ellipsis.

Let us clarify another conceptual issue on matter- vs. graviton- loop effects.
In the semi-classical limit only the matter fields are treated at the quantum level. This may seem to indicate that what's important for the trans-Planckian energy is the overall quantum effects - regardless of whether they come from matter or graviton fields - but not necessarily the quantum {\em gravitational} effects. This is not so. 
The loops of the matter fields introduce a cosmological constant term. Now one can consider the back reaction of the metric to the quantum-induced cosmological constant through the existence of the time-dependent solution. So strictly speaking, it is the quantum effects (regardless of whether they are the matter- or graviton- originated) plus the metric back reaction that are important for the trans-Planckian energy.  The fact that one considers the metric back reaction reflects that the metric is dynamical. Once one considers dynamical metric and matter quantum effect, there is no rationale to exclude the graviton loop effect, hence the relevance of the quantum gravitational effects. 
Related to this, the following can be said. The AMPS argument in \cite{Almheiri:2012rt} is based largely on the semi-classical framework but nevertheless leads to the trans-Planckian energy. Their argument certainly contains the {\em matter} quantum field-theoretic ingredient. The metric is perceived as dynamical and plays a dynamical role in the AMPS argument. By the same logic as above, the quantum {\em gravitational} ingredient is involved.

\vspace{.2in}
The following are the questions that can be answered by further extending the line of our recent research.\footnote{
Another more serious issue in the perturbative analysis is the long-known gauge-choice dependence. It is a more fundamental question \cite{Odintsov:1989gz,Odintsov:1991fk,Goncalves:2017jxq}.} 

\vspace{.1in}

The trans-Planckian energy results in a manner similar to that of \cite{Park:2017dib} where the scalar field scaling as $\sim \fr1{\k}$ led to the energy of order $\fr1{\k^2}$. Matter fields are present in both cases. It will be of some interests to explore the
question of whether or not a matter field is required for a trans-Planckian energy in general, especially in the context of the distorted black hole solutions of \cite{Moskalets:2014hoa} and \cite{Moskalets:2016uno}.

We believe that the trans-Planckian energy will typically occur in time-dependent situations. Even the time-independent cases could actually translate into time-dependent cases in a more realistic framework where one would consider an infalling {\em wave-packet} in the second-quantized Schrodinger framework. In that approach to which the present approach should be complementary, one would take $|vac>$ to be a certain type of wave-packet propagating in the background under consideration. The expectation value of the stress-energy tensor would be  computed with respect to the ``wave-packet vacuum." The onshell value of the Hamiltonian density will describe the spacetime-dependent energy density and the energy density around the packet will be time-dependent. This way, the time-dependence will be naturally built-in and we anticipate that the energy density will yield a trans-Planckian value around the packet as it approaches the horizon.

The course of our recent research repeatedly points to the importance of the boundary dynamics in a gravitational theory. 
In the present work, it was the non-vanishing boundary mode $\z_0^h(t)$ (more precisely, the horizon mode $\tilde{\z}_0^h(t)$) that led to the trans-Planckian energy.
More primarily, incorporation of various boundary conditions is necessary for correct identification of the whole Hilbert space of the theory \cite{Donnelly:2016auv}\cite{Freidel:2016bxd}. The widely-used Dirichlet boundary condition may well be of measure zero among all possible non-Dirichlet boundary conditions. We have analyzed the issue of the Dirichlet vs. non-Dirichlet boundary condition in detail in \cite{Park:2016vam}. It didn't appear possible to interpret the boundary condition of the quantum-level solution as a Neumann type. It might, however, be possible to interpret it as a Neumann-type up to peculiarities of an AdS spacetime. It will be of some interest to make this more accurate.

Closely tied with the boundary condition is the question of the stress-energy tensor. The definition of the stress-energy tensor itself has a long history of debates.  
Most of these debates were on the definition of the stress-energy tensor at the classical level \cite{Chen:1998aw,Kim:2013zha}. 
The quantization procedure poses additional subtleties.
One of the most serious issues should again be the one associated with the boundary terms and conditions.
A detailed analysis of the stress-energy tensor incorporating the works on the boundary terms such as \cite{Deruelle:2009zk} and \cite{Teimouri:2016ulk} should be performed.

\vspace{.1in}
We will report on the progress in some of these issues in the near future.




\newpage
\appendix

\renewcommand{\theequation}{A.\arabic{equation}}
\setcounter{equation}{0}

\section{Some details on geodesic equations}

Here we present some details on solving the geodesic equations.

\subsection{Schwarschild-Melvin case}

The coordinates $(r,\th)$ satisfy the following second order geodesic equations (see, e.g., \cite{Lim:2015oha} for the detailed study):
\bea
&&\ddot{r}+ \left(\frac{\partial _r\Lambda }{\Lambda } -\frac{f'}{2 f}\right) \dot{r}^2+\frac{1}{2} f  \left(f'+\frac{2 f \partial _r\Lambda }{\Lambda }\right)\dot{t}^2+\frac{f r   (r \partial _r\Lambda -\Lambda )}{\Lambda ^5}\sin ^2\theta  \dot{\phi }^2 \nn\\
&&  \hspace{2in}  -f  r\Big( 1 +\frac{r \partial _r\Lambda }{\Lambda }\Big)\dot{\theta }^2+\frac{2  \partial _{\theta }\Lambda }{\Lambda }\dot{\theta } \dot{r}
=0   \la{rdd} 
\eea
\bea
\hspace{-.2in}\ddot{\th}-\fr1{fr^2}\frac{\partial _{\theta }\Lambda }{ \Lambda  }  \dot{r}^2  +\fr{f}{r^2}\frac{ \partial _{\theta }\Lambda  }{\Lambda  }\dot{t}^2  +2 \left(\frac{1}{r}+\frac{\partial _r\Lambda }{\Lambda }\right) \dot{\theta } \dot{r}  +\frac{ \sin \theta  (\partial _{\theta }\Lambda  \sin \theta -\Lambda  \cos \theta )}{\Lambda ^5} \dot{\phi }^2 +\frac{ \partial _{\theta }\Lambda }{\Lambda }\dot{\theta }^2
=0  \la{thdd}\nn\\
\eea

Upon substituting $\dot{t}$ and $\dot{\f}$ into  \rf{vsq}, \rf{rdd} and \rf{thdd}, one gets
\bea
\frac{\Lambda ^2 }{f} \dot{r}^2 + \Lambda ^2 r^2 \dot{\theta }^2  -  \fr{E^2}{f \L^2}   + \fr{l^2 \L^2}{r^2\sin^2\th}      =s   
\la{vsq2} 
\eea
\bea
&&\ddot{r}+ \left(\frac{\partial _r\Lambda }{\Lambda } -\frac{f'}{2 f}\right) \dot{r}^2   -f  r\Big( (1 +\frac{r \partial _r\Lambda }{\Lambda }\Big)\dot{\theta }^2
  +\frac{2  \partial _{\theta }\Lambda }{\Lambda }\dot{\theta } \dot{r} \nn\\
&& +\frac{E^2}{2}  \fr{1}{\L^4} \left(\fr{f'}{f}+\frac{2  \partial _r\Lambda }{\Lambda }\right)  +  l^2\fr{ f}{r^3\sin^2\th} {   \Big(r \fr{\partial _r\Lambda}{\L} -1 \Big)} 
=0   \la{rdd2} 
\eea

\bea
&&\ddot{\th}-\fr1{fr^2}\frac{\partial _{\theta }\Lambda }{ \Lambda  }  \dot{r}^2 +\frac{ \partial _{\theta }\Lambda }{\Lambda }\dot{\theta }^2   +2 \left(\frac{1}{r}+\frac{\partial _r\Lambda }{\Lambda }\right) \dot{\theta } \dot{r}   \nn\\
&& + \fr{E^2}{f r^2} \frac{ \partial _{\theta }\Lambda  }{\Lambda^5  }   +  \fr{l^2 }{r^4\sin^3\th} {   \Big(\sin \theta \fr{\partial _{\theta }\Lambda}{\L}  -  \cos \theta \Big)}
=0  \la{thdd2}
\eea
With $\th=\pi/2$, eq.\rf{thdd2} is satisfied
and eq.\rf{vsq2} can be solved and one gets for $\ddot{r}$:
\bea
\ddot{r}=-\fr{4E^2B^2r}{(1+B^2r^2)^5}+l^2\fr{(r-3M)}{r^4}+s\fr{(M+5MB^2r^2-2B^2r^3)}{r^4(1+B^2r^2)^3}
\eea

\subsection{generalized Schwarschild-Melvin case}

One can verify that $E=-g_{vv} \dot{v}-g_{vr} \dot{r}-g_{v\th}\dot{\th}$ and $l=g_{\vf\vf} \dot{\vf}$ are the first integrals (energy and angular momentum) for the $\vf$ and $v$ geodesic equations
\[
\fr{d^2 \vf}{du^2}+2\left(\fr1{r}+\fr{\csc^2\th}{2r^3}\left(r\pa_r \gt_{\vf\vf}-2\gt_{\vf\vf}\right)\right)\fr{dr}{du}\fr{d\vf}{du}
\]
\be
+2\left(\cot \th+\fr{\csc^2\th}{2r^2}\left(\pa_\th \gt_{\vf\vf}-2\cot\th \gt_{\vf\vf}\right)\right)\fr{d\th}{du}\fr{d\vf}{du}=0,
\la{feq1}
\ee
\[
\fr{d^2 v}{du^2}+\left(\fr{f'}2-\fr12 \pa_r \gt_{vv} \right)\left(\fr{dv}{du}\right)^2-\pa_r \gt_{v\th} \fr{dv}{du}\fr{d\th}{du}
\]
\be
-\left(r+\fr12 \pa_r \gt_{\th\th}\right)\left(\fr{d\th}{du}\right)^2-\left(r\sin^2\th+\fr12 \pa_r \gt_{\vf\vf} \right)\left(\fr{d\vf}{du}\right)^2=0.
\la{veq1}
\ee
for the metric \rf{gvfvfpert}.
For instance, one can check that $u$-differentiation of
\be
\dot{\vf}=\fr{l_0}{r^2 \sin^2\th+\tilde{g}_{\vf\vf}}=\fr{l_0}{r^2 \sin^2 \th}\left(1-\fr{\tilde{g}_{\vf\vf}}{r^2 \sin^2\th}\right)+\cO(\tilde{g}^2_{\vf\vf})
\la{dotvf1st}
\ee
results in eq.\rf{feq1}. The details are as follows. By taking the second derivative one gets
\[
\ddot{\vf}=-\fr2{r}\left(\fr{l_0}{r^2 \sin^2\th}\,\dot{r}+\fr{l_0}{r^2 \sin^2\th}\,\cot{\th} \,r \dot{\th}\right)
\]
\[
+\fr{1}{r^3\sin^2\th}\left(4\,\fr{l_0}{r^2\sin^2\th}\,\tilde{g}_{\vf\vf} \dot{r}+4\,\fr{l_0}{r^2\sin^2\th}\,\cot\th\,\tilde{g}_{\vf\vf} r\dot{\th} \right.
\]
\be
\left. -\fr{l_0}{r^2\sin^2\th}\,r\dot{\th}\pa_\th\, \tilde{g}_{\vf\vf}-\fr{l_0}{r^2\sin^2\th}\,r\dot{r}\pa_r\, \tilde{g}_{\vf\vf} \right)
\la{ddotvf1st}
\ee
One arrives at eq. \rf{feq1} by combining \rf{dotvf1st} and \rf{ddotvf1st}.

\newpage

\end{document}